\documentclass[twocolumn,showpacs,preprintnumbers,amsmath,amssymb]{revtex4}
 \usepackage{graphicx}
 \usepackage{dcolumn}
 \usepackage{bm}
 \begin{document}

 \title{Anti-correlation between energy-gap and phonon energy for cuprate Bi2212 superconductor}

 \author{Wei Fan}
 \affiliation{Key Laboratory of Materials Physics, Institute of Solid State Physics, Chinese
 Academy of Sciences, 230031-Hefei, People's Republic of China}
 \date{\today}

 \begin{abstract}
 Using electron-phonon mechanism, we explains very well
 the spatial anti-correlation between the energy-gap and the energy of phonon mode
 for cuprate superconductor found in $d^{2}I/dV^{2}$ spectrum
 by STM measurements [Jinho Lee, et.al Nature {\bf 442},546 (2006)],
 which is is the direct effect of a relationship
 $M\langle\omega^{2}\rangle\lambda=const$.
 We calculate T$_{C}$ maps on $\lambda-\Omega_{P}$ plane which
 helps us understanding the relation $M\langle\omega^{2}\rangle\lambda=const$
 and the superconductivity of superconductors.
 \end{abstract}
  \maketitle


 Scanning Tunneling Microscopy/Spectroscopy (STM/S) has been used to explore
 surface structures and electronic structures of superconductors~\cite{Fischer1}.
 The electronic inhomogeneity such as the inhomogeneously distributed energy-gaps
 and local densities of states has been found in High Temperature
 Superconductor (HTSC)~\cite{Pan1,Lee1}.
 The intrinsic inhomogeneity even exists when translational invariance of
 lattice structure preserves. If total momentum of cooper pair isn't equal zero,
 in real-coordinate space the distribution of cooper pairs is inhomogeneous.
 The Fulde-Ferrell-Larkin-Ovchinnikov (FFLO) phase can be realized by
 applying magnetic field or stress field to superconductors.

 An interesting property of electronic inhomogeneity of superconductor by
 the observations of STM/S for HTSC Bi2212 is the spatial anti-correlations of
 energy-gap and energy of boson mode~\cite{Lee1}. The understanding of
 origin of the anti-correlation is very useful to understand the microscopic mechanism
 of superconductivity for HTSC materials. The boson mode for Bi2212 has been
 identified as phonon mode due to the isotope effects.

 The out-of-plane buckling modes, in-plane breathing modes and the phonon modes related apical
 oxygen atoms are expected to contribute to superconductivity of HTSC materials.
 The out-of-plane buckling modes coupling with the electrons near anti-node region have small
 momentum transfer and significant band re-normalization effects near transition temperatures
 which had been found in ARPES spectrum and theoretical calculation~\cite{Devereaux1}.

 In this paper, we preform the standard calculations of Nambu-Eliashberg strong coupling
 theory and focus on an important relation $M\langle\omega^{2}\rangle\lambda=\eta=const$.
 The superconductor parameter $\eta=\rho(0)\langle J^{2}\rangle$, the product of the electronic
 properties $\rho(0)$ the density of state at Fermi energy and the electron-phonon matrix elements
 $J$, characterizes the chemical environments of atoms and almost keeps as a constant
 against the simple structural changes such as the isotope substitution.
 We assume the electron-phonon mechanism is pairing mechanism for
 superconductor Bi2212 in this paper because of the isotope effects and identification of
 phonon mode in ARPES spectrum~\cite{Damascelli1}, STM/S experiments~\cite{Lee1}.
 In terms of the relation $M\langle\omega^{2}\rangle\lambda=\eta$, we explain the anti-correlation between
 energy-gap $\Delta$ and phonon energy $\Omega_{P}$. We also plot two T$_{C}$ maps of
 transition temperature on $\lambda-\Omega_{P}$ plane, which are very useful to
 understand superconductivity of conventional and unconventional superconductors.

 The theoretical methods used in this paper are (1) the real-energy method by having analytically
 continued the imaginary Eliashberg Equation to real axis~\cite{Scalapino1,Holcomb1}
 and (2) the imaginary-energy Matsubara method~\cite{Allen1} of strong coupling superconducting theory
 under isotropic approximation. After having resolved isotropic energy-gap equation,
 we can obtained the anisotropic energy-gap by directly multiplying
 isotropic energy-gap $\Delta(\omega)$ with the anisotropic function such as for
 the d-wave the anisotropic energy-gap $\Delta(\omega)\cos(2\theta)$ and then averaging
 it over $\theta$ angle. The anisotropy of energy-gap comes from the
 anisotropic electron-phonon interaction~\cite{Devereaux1,Sandvik1}.
 The real-energy method gives the results of densities of states, energy gaps and T$_{C}$.
 The advantage of imaginary methods is that we can quickly obtain the T$_{C}$ compared
 with the real-energy methods.

 The densities of state of phonon is calculated by following formula~\cite{Scalapino1}
 \begin{eqnarray}\label{eq1}
 F(\omega)=\left\{
 \begin{tabular}{cc}
  $\frac{c}{(\omega-\Omega_{p})^{2}+(\omega_{2})^{2}}
  -\frac{c}{(\omega_{3})^{2}+(\omega_{2})^{2}}$, &
  $|\omega-\Omega_{p}|<\omega_{3}$ \\
  0 & others,
 \end{tabular}
 \right.
 \end{eqnarray}
 \noindent where $\Omega_{p}$ is the energy of phonon mode, $\omega_{2}$ the half-width
 of peak of phonon mode and $\omega_{3}^\lambda=2\omega_{2}$.
 The effective electron-phonon interaction $\alpha^{2}_{y}(\nu)$ is defined by
 \begin{eqnarray}\label{eq2}
 \alpha^{2}_{y}(\omega)F_{y}(\omega)=\frac{1}{W}
 \int_{S}\frac{d^{d-1}p}{v_{F}}
 \int_{S'}\frac{d^{d-1}p'}{v_{F'}} B_{y}(p-p',\omega)|D^{e}_{p-p',y}|^{2}.
 \end{eqnarray}
 \noindent where $W=1/(2\pi)^{d-1}\int_{S}d^{d-1}p/v_{F}$ and $v^{F}$ Fermi velocity.
 In 2-dimensional limit the coupling between different
 layers being ignored, we can solve the Eliashberg equation for
 2-dimensional systems. Compared with 3-dimensional system, the energy-gap equation
 keeps unchanged and all changes are included in the Eliashberg function $\alpha^{2}F(\omega)$.
 Complete treatments of anisotropic effects need the k-dependent
 energy-gap function $\Delta (k,\omega)$ and re-normalized factors $Z(k,\omega)$.

 \begin{figure}
 \includegraphics[width=0.40\textwidth]{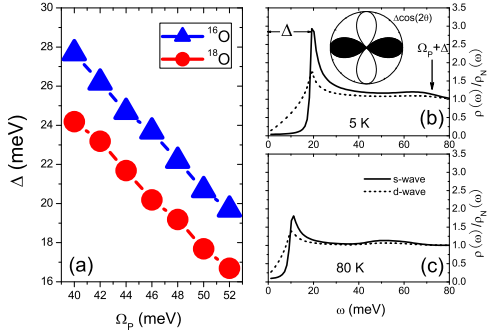}
 \caption{\label{fig1}
 (a). The anti-correlation between energy-gap $\Delta$ and the phonon energy $\Omega_{P}$
  and corresponding isotope effects for $^{16}$O substituted by
  $^{18}$O. (b,c). The s-wave and d-wave densities of states at 5 K and 80 K.
  The definition of energy gap $\Delta$ by densities of states (DOS)
  and energy gap with d-wave symmetry in (b).}
 \end{figure}

 We can write the parameter of electron-phonon interaction $\lambda=\sum_{y}\lambda_{y}$,
 where parameter $\lambda_{y}=2\int_{0}^{\infty}d\omega\alpha^{2}_{y}F_{y}(\omega)/\omega$
 is the contribution of $y$ mode to total $\lambda$.
 The moments $\langle\omega^{n}\rangle$ of the distribution function
 $(2/\lambda)\sum_{y}2\alpha^{2}_{y}F_{y}(\omega)/\omega$ are defined as
 $\langle\omega^{n}\rangle=2/\lambda\sum_{y}\int^{\infty}_{0}d\omega\alpha^{2}_{y}F_{y}(\omega)\omega^{n-1}$.
 In this work we assume that $\alpha_{y}$ is constant around each
 peak in phonon spectrum. Under this approximation, we only consider
 $\alpha^{2}F(\omega)$ as a whole, the so-called Eliashberg function with above Lorentz shape.
 In practice, by modifying the height of peak of Eliashberg function $\alpha^{2}F(\omega)$,
 we can control the change of the parameter $\lambda$ of electron-phonon interaction.
 The densities of states of electron are obtained from the formula
 $\rho(\omega)=\rho_{N}(\omega)Re(\omega/\sqrt{\omega^{2}-\Delta(\omega)^{2}})$.
 If the anisotropy of energy gap is considered such as with the d-wave symmetry,
 the density of state $\rho(\omega)=\rho_{N}(\omega)\langle
 Re(\omega/\sqrt{\omega^{2}-(\Delta(\omega)\cos(\theta))^{2}})\rangle_{\theta}$,
 with the $\langle\ldots\rangle_{\theta}$ describing the average over $\theta$.

 We choose the phonon spectrum of Eq.~(\ref{eq1}) with phonon-energy about
 $\Omega_{P}$=52 meV, which is the mean phonon-energy observed in the tunneling spectrum
 of Bi2212~\cite{Lee1}. The $\Omega_{P}$ distributes from 30 meV to 70 meV and
 within the range from about 36 meV the B$_{1g}$ phonon coupling to electrons at antinode
 region and to 70 meV the in-plane breathing mode coupling with electrons at node region
 based on the observation of ARPES spectrum~\cite{Damascelli1}.

 If the parameter $\lambda$ of electron-phonon interaction is 2.0 and the
 Coulomb pseudo-potential $\mu^{*}$=0.24, the transition temperature
 T$_{C}$ is about 90K close the experimental values. From STM experiments of Bi2212,
 the energies of phonon have a distribution with width
 about $2\omega_{2}$=16 meV. The energy gap $\Delta$ and the $\lambda$ the parameter of
 electron-phonon interaction should have distributed with their widthes respectively.
 If we consider a material with multi-species atoms, total electron-phonon interaction parameter
 should be expressed as $\lambda=\sum_{m}\lambda_{m}$ with
 $\lambda_{m}=\eta_{m}/M_{m}\langle\omega_{m}^{2}\rangle$
 the electron-phonon interaction parameters for every species atoms.
 If the phonon modes correlated with the vibrations of oxygen atoms
 have most significant contributions to superconductivity,
 we can get simple relation $M\langle\omega^{2}\rangle\lambda=\eta=const$.
 Additionally, we assume it is correct for isotope substitution, when mass of oxygen atom
 increases from 16 a.u to 18 a.u.  Both $\lambda$ and $\langle\omega^{2}\rangle$ should be changed
 to keep $M\langle\omega^{2}\rangle\lambda=\eta=const$.

 The Fig.~(\ref{fig1}) plots the $\Delta-\Omega$ relation for the samples with $^{16}$O and $^{18}$O.
 Certainly, the products $M\langle\omega^{2}\rangle\lambda$ are all the same for all the
 points in the Fig.~(\ref{fig1})(a). The curve for $^{18}$O are beneath the curve
 for $^{16}$O.  The very similar result is found in Fig.5(a) in Ref.~\cite{Lee1}.
 The energy-gaps decrease with increasing phonon energy $\Omega_{P}$ because the
 parameters of electron-phonon interaction increase with decreasing phonon energy
 because $M\langle\omega^{2}\rangle\lambda=const$. In a sample of superconductor, the
 in-homogeneously distributed phonon-energies $\Omega_{P}(r)$ lead to the in-homogeneously
 distributed energy-gap $\Delta(\omega,r)$, where $\Omega_{P}$ is large, $\Delta(\omega,r)$
 small. The vertical shift of $\Delta-\Omega_{P}$ curve downward to smaller value for larger mass
 is due to the increasing parameter of electron-phonon interaction with
 decreasing Mass because of the relation $M\langle\omega^{2}\rangle\lambda=const$.
 However, the energy gaps $\Delta$ in this work are only about half of values in
 Ref.~\cite{Lee1} and the vertical shift of 4.0 meV in $\Delta$ smaller than 5.6 meV in
 Ref.~\cite{Lee1}. The reason is that the energy gaps in this
 paper are directly correlated to the formation of coherent cooper pairs and the energy gaps in
 Ref.~\cite{Lee1} are the mixing of superconducting gap and pseudo-gap of spin
 fluctuations. The values of superconducting energy gap is about from 23 meV
 larger than the values 19-20 meV in this work. The effects of the relationship
 $M\langle\omega^{2}\rangle\lambda=const$ had already been discussed in Ref.\cite{McMillan1,Allen1}.

 \begin{figure}
 \includegraphics[width=0.40\textwidth]{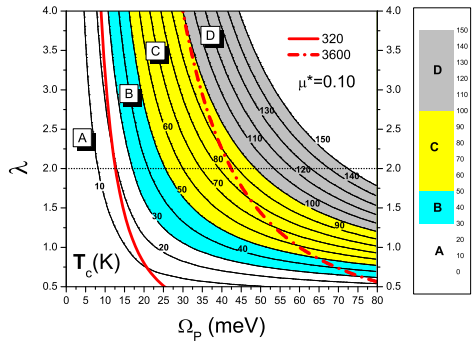}
 \caption{\label{fig2}
  The T$_{C}$ map on $\lambda-\Omega_{p}$ plane with $\mu^{*}$=0.10 obtained from
  imaginary-energy Matsubara method. The bold solid line and dash-dot line
  are the curves $\lambda\Omega^{2}_{P}$=320 and 3600 (meV)$^{2}$.
  The region $A$ includes the contour lines from 0 K to 30K, $B$ from 30 K to 50 K,
  $C$ from 50 K to 100 K and $D$ from 100 K to 150 K.}
 \end{figure}

  \begin{figure}
 \includegraphics[width=0.40\textwidth]{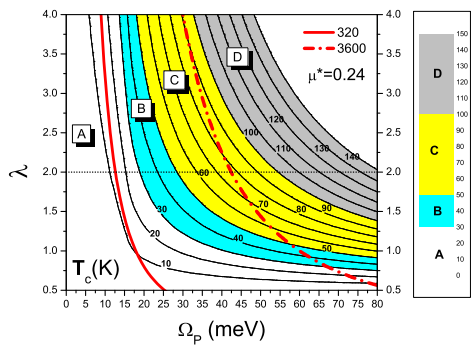}
 \caption{\label{fig3}
  The T$_{C}$ map on $\lambda-\Omega_{p}$ plane with $\mu^{*}$=0.24 obtained from
  imaginary-energy Matsubara method. The bold solid line and dash-dot line
  are the curves $\lambda\Omega^{2}_{P}$=320 and 3600 (meV)$^{2}$}
 \end{figure}

 A T$_{C}$ map on the parameters space $\lambda-\Omega-\mu^{*}$ is very
 helpful to understand the superconductivity with electron-phonon mechanism.
 The important role of the relationship $M\langle\omega^{2}\rangle\lambda=const$
 can be shown clearly on the T$_{C}$ map. In this paper we provide T$_{C}$ maps on 2-dimensional
 $\lambda-\Omega$ plane with 50$\times$ 50 mesh-grid with parameters $\lambda$ from 0.5 to
 4.0 and $\Omega_{P}$ from 0 meV to 80 meV. We plot two different T$_{C}$ maps with the
 Coulomb pseudo-potential $\mu^{*}$=0.1 Fig.~(\ref{fig2}) and 0.24 Fig.~(\ref{fig3}).
 For every mesh points we can calculate a T$_{C}$ using imaginary-energy Matsubara methods.
 An important point is that, for high energy phonon mode which is the vibrations of
 relatively light oxygen atoms, only intermediate $\lambda (\le 2.0)$ or
 intermediate electron-phonon coupling interaction can guarantee the values of
 T$_{C}$ within the range of HTSC materials.
 As we expect that for large Coulomb pseudo-potential $\mu^{*}$=0.24 the
 larger parameters $\lambda$ of electron-phonon interaction are needed
 to obtained the same T$_{C}$ for the general values $\mu^{*}$=0.1.

 In two figures we also plot two curves of $\Omega^{2}_{P}\lambda$=320 and 3600 (meV)$^{2}$.
 We have been ignored the changes of mass of atoms in the relation
 $M\langle\omega^{2}\rangle\lambda=const$. We can see that the two curves cross
 smaller number contours lines when $\lambda>2.0$ with small phonon energies.
 This fact indicates that product $\langle\omega^{2}\rangle\lambda$ is
 an important parameters characterized T$_{C}$. Additionally, the curve for
 $\Omega^{2}_{P}\lambda$= 3600 (meV)$^{2}$ enters the region with higher transition
 temperature compared with the curve for $\Omega^{2}_{P}\lambda$= 320 (meV)$^{2}$.
 Although the range of $\lambda > 2.0$ is unfavorable superconductivity due to
 the lattice instability, the product is still good parameters characterized T$_{C}$.
 When superconductor under small structural modifications or isotope substitution,
 the corresponding parameters $\lambda$ and $\Omega_{P}$ move alone these curves
 because the chemical environments keep almost unchanged.

 For high-energy phonon, the smaller change of $\lambda$ can induce large change of T$_{C}$
 because the curve $\Omega^{2}_{P}\lambda$=3600 cross more contour-lines than for low-energy phonon.
 So the superconductivity correlated with high-energy phonon is structural sensitivity.
 The energy of phonon plays more important role for high-temperature superconductor.
 Because the atoms with smaller mass have larger frequency of vibrations, almost all
 superconducting materials with high transition temperature contain light atoms such as
 Carbon atoms in fullerides and oxygen atom  in HTSC materials. However these materials
 face the problem of instability of superconductivity induced by the structural instability.

 \begin{table}
 \caption{\label{table1} Transition temperatures T$_{C}$ calculated
 from the real-energy method
 T$^{Re}_{C}$ and imaginary-energy Matsubara method T$^{Im}_{C}$.
 The relative errors $e=(T^{Re}_{C}-T^{Im}_{C})/T^{av}_{C}$, where $T^{av}_{C}=(T^{Re}_{C}+T^{Im}_{C})/2$ }
 \begin{center}
 \begin{ruledtabular}
 \begin{tabular}{ccccccc}
 \hline
 & $\Omega_{p} (meV)$ & $\mu^{*}$  &  $\lambda$ &  $T^{Re}_{C} (K)$ & $T^{Im}_{C} (K)$ &  $e$ \\
 \hline
 (1)&25 &0.10 &0.8 &14   &15  &0.069 \\
 (2)&25 &0.10 &1.5 &33   &37  &0.114 \\
 (3)&25 &0.24 &1.8 &34   &39  &0.137 \\
 (4)&25 &0.24 &2.0 &38   &43  &0.123 \\
 (5)&52 &0.10 &0.8 &33   &35  &0.059 \\
 (6)&52 &0.10 &1.5 &79   &82  &0.037 \\
 (7)&52 &0.24 &1.8 &79   &86  &0.085 \\
 (8)&52 &0.24 &2.0 &90   &95  &0.054 \\
 \end{tabular}
 \end{ruledtabular}
 \end{center}
 \end{table}

 The T$_{C}$ values obtained using two different methods (real-energy and imaginary methods)
 only have small differences with relative errors $e<0.14$ if the high-energy cutoff is larger
 than 1000 meV and the low-energy cutoff is smaller than 1.20 meV.
 In the real-energy method, the energy cutoff is about 1000 meV at high energy and
 1.20 meV at low energy. In imaginary method, we use about 200 Matsubara energies to
 resolve the Eliashbeg equation. From the table~(\ref{table1}) the larger relative errors
 happen at strong coupling regime with larger Coulomb pseudo-potentials and however
 with smaller phonon energies such as the case(2,3,4). With increasing the energies
 of phonon from 25 meV to 52 meV, the relative errors will smaller than 0.1.
 The absolute values of errors from 5K to 7K for the case (6,7,8) are small for
 high transition temperature T$_{C}$ however large for the case (3,4) with relative
 low transition temperatures. From above comparisons, we can say that the T$_{C}$
 map in Fig.~(\ref{fig2}) is helpfully although there are some uncertain from
 the numerical calculations.

 In summary, we find the anti-correlation between energy-gap $\Delta$ and phonon energy
 $\Omega_{P}$ is deduced from the simple relation $M\langle\omega^{2}\rangle\lambda=const$
 in strong coupling theory. We also plot two T$_{C}$ maps in $\lambda-\Omega_{P}$ to
 understand the above relation and the superconductivity. This work is supported by Knowledge
 Innovation Program of Hefei Institutes of Physical Sciences, Chinese Academy of Sciences,
 and in part by National Science Foundation of China for CAS projects.

\end{document}